\documentclass[prl,twocolumn,showpacs,preprintnumbers,amsmath,amssymb]{revtex4}
\usepackage{graphicx}
\usepackage{dcolumn}
\usepackage{amssymb}
\usepackage{amsmath}

\begin{document}

\title{Defect Energy Levels in Density Functional Calculations: Alignment and Band Gap Problem}

\author{Audrius Alkauskas}
\author{Peter Broqvist}
\author{Alfredo Pasquarello}
\affiliation{Ecole Polytechnique F\'ed\'erale de Lausanne
(EPFL), Institute of Theoretical Physics, CH-1015 Lausanne, Switzerland}
\affiliation{Institut Romand de Recherche Num\'{e}rique en Physique des Mat\'{e}riaux (IRRMA),
CH-1015 Lausanne, Switzerland}
\date{\today}
\begin{abstract}
For materials of varying band gap, we compare energy levels of atomically localized defects 
calculated within a semilocal and a hybrid density-functional scheme.  Since the latter scheme partially 
relieves the band gap problem, our study describes how calculated
defect levels shift when the band gap approaches the experimental value.
When suitably aligned, defect levels obtained from 
total-energy differences correspond closely, showing average shifts 
of at most 0.2 eV irrespective of band gap. Systematic deviations from  
ideal alignment increase with the extent of the defect wave function. A guideline  
for comparing calculated and experimental defect levels is provided.
\end{abstract}

\maketitle
Semilocal approximations to density functional theory, such as the local density approximation (LDA) and the
generalized-gradient approximation, have proved extremely valuable to investigate energetic, atomistic,
and magnetic properties of defects in solids \cite{VanDeWalle_JAP_2004}.
However, these approximations have been much less successful in locating charge transition levels in the
band gap, because of the well known band gap problem from which they suffer.
As a result, a direct comparison between calculated and experimental energy levels remains ambiguous.
Furthermore, the determination of equilibrium densities of intrinsic defects and charge carriers
is hindered \cite{Lany_PRL_2007}.
Therefore, considerable efforts have been deployed in the study of defects to address the band-gap problem
going beyond semilocal approximations to density functional theory.
Many-body perturbation theory in the $GW$ approximation is the method of choice for
calculating defect levels \cite{Ismail-Beigi_PRL_2005},
but remains computationally demanding and therefore limited to small-size systems.
Several practical routes have also been proposed, such as the scissor-operator scheme,
the marker method \cite{marker},
the LDA+$U$ method \cite{Janotti_PRB_2007}, the use of adapted pseudopotentials \cite{Stampfl_PRB_2000}, 
and the application of ad-hoc extrapolation
schemes \cite{Zhang_PRB_2001}. However, the general applicability of these approaches is unclear.
More recently, hybrid density functionals have become increasingly popular for addressing defect energy levels
\cite{Knaup_PRB_2005}.
These functionals incorporate a fraction of Hartree-Fock exchange, leading to higher
accuracy \cite{Curtiss_JCP_1997} and improved band gaps \cite{Muscat_CPL_2001} compared
to semilocal functionals.

In this work, we carry out a comparative study between defect energy levels calculated with semilocal and
hybrid density functionals to determine their shifts as the description of the band gap improves.
We aim at gaining insight into how calculated and measured defect levels should be compared when the adopted
theoretical scheme is subject to the band gap problem.
For this purpose, we considered materials covering a large range of band gaps and selected defect levels 
spanning large portions of their band gaps.
Our study shows that charge transition levels obtained with semilocal and hybrid density functionals
correspond closely, provided a suitable alignment scheme is adopted. 
As the band gap decreases, systematic 
deviations from ideal alignment are found to increase with the extent of the defect wave function.  

The semilocal density-functional calculations were performed within the generalized gradient
approximation proposed by Perdew, Burke, and Ernzerhof (PBE) \cite{PBE_PRL_1996}.
We used a hybrid density functional, denoted PBE0, which is obtained from the latter by
replacing 25\%\ of the PBE exchange energy by Hartree-Fock exchange \cite{PBE0}.
We used a scheme based on plane-wave basis sets and normconserving 
pseudopotentials.
The pseudopotentials were generated at the semilocal level and used in all calculations.
The plane-wave basis set was defined by an energy cutoff of 70 Ry. The Brillouin
zones of our supercells were sampled at the $\Gamma$ point, but primitive cells with
a converged $k$-point sampling were used for the determination of the bulk band edges. 
We took care of the integrable divergence of the Hartree-Fock exchange term  
\cite{Gygi_PRB_1986}.
Structural relaxations were carried out at the semilocal level \cite{note1}.
Our calculations were performed with the codes \textsc{q}uantum-\textsc{espresso} 
\cite{QE} and \textsc{cpmd} \cite{CPMD}.

\begin{table}
\caption{Calculated and experimental band gaps (in eV).}
\begin{ruledtabular}
\begin{tabular}{l c c c c}
          &  Si    &  SiC   &  HfO$_2$  &  SiO$_2$              \\ \hline
Semilocal &  0.6   &  2.2   &  4.3      &  5.8                  \\
Hybrid    &  1.8   &  3.9   &  6.7      &  8.3                  \\
Expt.     &  1.2   &  3.3   &  5.9      &  8.9                  \\
\end{tabular}
\label{tab1}
\end{ruledtabular}
\end{table}

We considered defects in four different materials of varying 
band gap: Si, SiC, HfO$_2$, and SiO$_2$.
The band gaps of these materials calculated at the semilocal level severely underestimate
the experimental values (Table \ref{tab1}), as usual for semilocal density functional 
schemes. As shown in Table \ref{tab1}, the hybrid scheme systematically gives 
larger band gaps, generally leading to a better agreement with experiment. 
For silicon, we adopted a cubic simulation cell of 64 atoms and considered 
the following defects with their relative charge states: the Si vacancy 
($+$2, $+$1, 0, $-$1, $-$2), the Si self-interstitial ($+$1, 0), the 
substitutional O (0, $-$1), the C interstitial ($+$1, 0, $-$1), and 
the P/Si-vacancy complex ($+$1, 0, $-$1). For SiC, we modeled the 4$H$ polytype 
using an orthorhombic cell containing 96 atoms. Considered defects include
the C vacancy ($+$2, $+$1, 0, $-$1, $-$2), the Si vacancy ($+$2, $+$1, 0, $-$1, $-$2),
and the complex consisting of C substitutional to Si next to a C vacancy 
($+$2, $+$1, 0, $-$1, $-$2).
For HfO$_2$, we took the monoclinic structure and used a supercell containing 96 atoms.
We considered the O vacancy ($+2$, $-1$, $0$, $+1$ and $-2$) and the O 
interstitial ($0$, $-$1, and $-2$).  
For SiO$_2$, we modeled $\alpha$-quartz with an orthorhombic cell containing 
72 atoms and considered the interstitial H ($+$1, $0$, $-$1), the Si-Si dimer
bond ($+$1, $0$), the puckered O vacancy ($+$1, $0$), the H 
bridge -Si-H-Si- ($+$1, $0$, $-$1), the substitutional N ($0$, $-$1),
and the interstitial O$_2$ ($0$, $-$1).
All defect states studied are atomically localized.

\begin{figure}
\includegraphics[width=6.2cm]{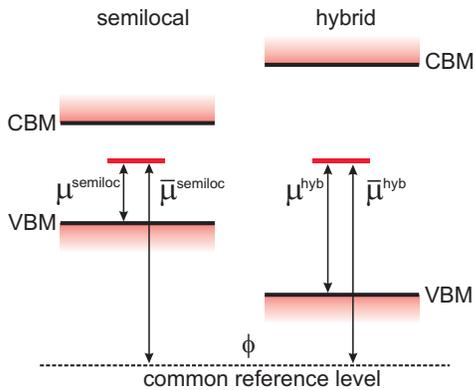}
\caption{(Color online) Schematic illustration of the alignment between energy levels obtained with a semilocal and a
hybrid density functional. The charge transition levels ${\mu}$
and $\overline{\mu}$ are referred to the respective valence band maxima
(VBM) and to a common reference level, respectively. 
The conduction band minima (CBM) are also shown.}
\label{fig1}
\end{figure}
 
The formation energy of a defect in its charge state $q$ can be expressed in terms of the electron
chemical potential $\mu$ referred to the valence band maximum $\varepsilon_\text{v}$ \cite{VanDeWalle_JAP_2004}:
\begin{equation}
E_{\text{f}}^{q}(\mu)=E^{q}_{\text{tot}}-E^{\text{bulk}}_{\text{tot}}-\sum_{\alpha}n_{\alpha}\eta_{\alpha}+q(\mu+\varepsilon_\text{v}),
\label{formation}
\end{equation}
where $E^{q}_{\text{tot}}$ is the total energy of the defect system, $E^{\text{bulk}}_{\text{tot}}$ the total
energy of the unperturbed system, $n_{\alpha}$ the number of extra atoms of species $\alpha$ needed to
create the defect, and $\eta_{\alpha}$ the corresponding atomic chemical potential.
Charge transition levels correspond to specific values of the electron chemical potential for
which two charge states have equal formation energies.
We considered both thermodynamic and vertical charge transition levels.

To compare defect levels in semilocal and hybrid density-functional schemes,
it is necessary to use a common reference level $\phi$ external to the electronic system,
i.e.\ defined on the basis of the nuclear potentials:
$ \overline{\mu}=\mu +\varepsilon_\text{v}-\phi$.
This alignment scheme is illustrated in Fig.\ \ref{fig1}. In our formulation,
we used the same pseudopotentials in the semilocal and hybrid calculations and trivially
achieved such an alignment by taking $\phi$ as the cell average of the local potential
originating from the ionic pseudopotentials.
However, we note that our formulation does not imply any loss of generality
and that a proper alignment can also be enforced when different pseudopotentials are used.

\begin{figure}
\includegraphics[width=8.5cm]{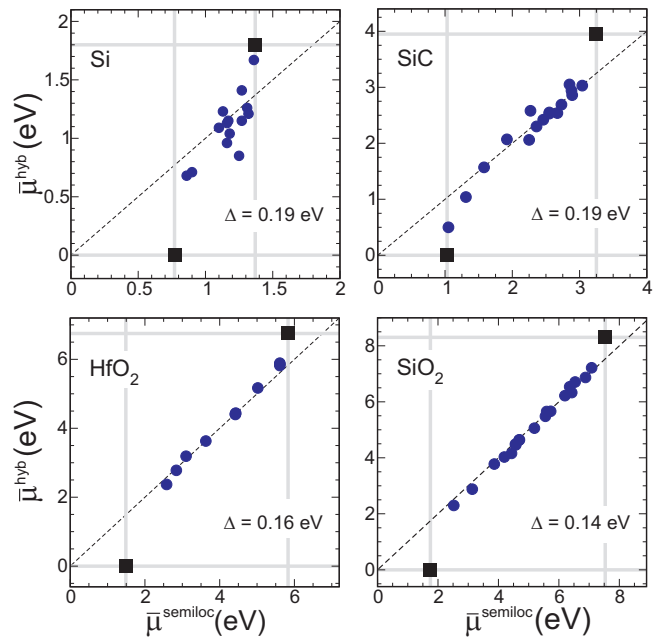}
\caption{(Color online) Comparison between charge transition levels calculated with the semilocal
($\overline{\mu}^\text{semiloc}$) and hybrid ($\overline{\mu}^\text{hyb}$) functionals
for a variety of defects in Si, SiC, HfO$_2$, and SiO$_2$.  The energy levels
corresponding to the valence band maximum (VBM) and conduction band minimum (CBM) are also
shown (squares). All energies are referred to a common reference level ${\phi}$
(see text), shifted to coincide with the VBM in the hybrid scheme for convenience. For each material, 
$\Delta$ is the r.m.s. error with respect to the ideal alignment (dashed).}
\label{fig2}
\end{figure}

%
We calculated charge transition levels for the selected set of defects in Si, SiC, HfO$_2$, 
and SiO$_2$ within both the semilocal and hybrid schemes. 
Charge transition levels obtained in either scheme were then aligned with respect to the
common reference level ${\phi}$ and reported in Fig.~\ref{fig2}.
For each material, our results show that the defect levels calculated in the semilocal and hybrid schemes
{\it differ on average by at most 0.2 eV} when aligned in this way, despite the significantly
larger concomitant variations observed for the band gaps.
Since average shifts are similar in the four cases studied, the identified alignment 
is more impressive for large band gap materials where these shifts are small with respect 
to the band gap. Indeed, the average relative shift is only 2\% for SiO$_2$, 
but increases to about 17\% for Si. In particular, these results indicate that differences 
between charge transition levels are already well described at the semilocal level, lending support 
to alignment schemes in which the defect levels are anchored to experimental marker 
levels given by well characterized defects \cite{marker,Schultz_PRL_2006}.
In this respect, a key result of our work is that the defect levels are positioned on an 
energy scale unaffected by band gap renormalization.

\begin{figure}
\includegraphics[width=8.0cm]{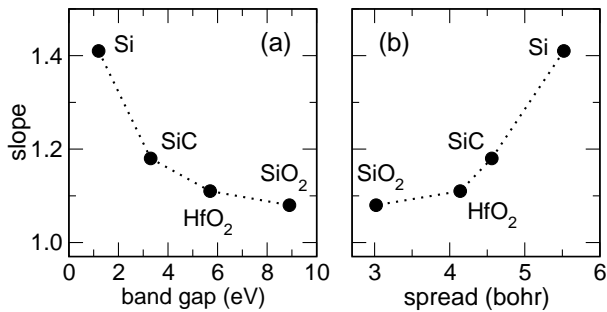}
\caption{   
Optimal slopes derived from linear regressions of the data in Fig.\ \protect\ref{fig2} as a 
function of (a) experimental band gap and (b) average spread of the defect wave functions.
}
\label{fig3}
\end{figure}

To reveal systematic deviations with respect to the ideal alignment, we carried out linear 
regressions of the available data deriving optimal slopes [Fig.~\ref{fig3}(a)]. Ideal alignment
corresponds to unitary slope and is best illustrated for SiO$_2$ (slope of 1.08).  
When the band gap decreases, the optimal slope is found to increase indicating that defects 
in the upper part of the band gap tend to follow the conduction band, 
while defects in the lower part of the band gap tend to follow the valence band. 
This tendency is most pronounced for Si (slope of 1.4).  

To provide a rationale for the obtained results, we first discard effects which affect the defect levels
in a minor way. In all the cases studied, irrespective of band gap, the electron wave functions were 
found to be very similar in the semilocal and hybrid schemes. Their effect on defect levels can be quantified 
by calculating total energies at the hybrid level using electron wave functions optimized in the semilocal scheme 
without alowing for electron relaxation \cite{Alkauskas_PB_2007}. In this way, we inferred that the 
differences between the defect levels calculated in the two schemes do not arise from variations of the 
electron wave functions.
Since the observed trends also hold for the subset of vertical transitions, we discard effects associated to 
differences in structural relaxation energies.
The Slater transition-state approximation then allows us to focus on the single defect eigenstate
rather than on the full manifold of occupied states \cite{Slater}. This approximation expresses the 
relevant total-energy differences by the energy eigenvalue of the defect state at half filling,
$\overline{\mu} \cong \langle \psi_{\rm D} | \cal{H}| \psi_{\rm D} \rangle  - \phi$,
and gives accurate charge transition levels in both schemes \cite{Alkauskas_PB_2007}.

In the Slater approximation, the difference in charge transition levels can then be expressed as 
\begin{equation}
\overline{\mu}^{\text{hyb}}-\overline{\mu}^{\text{semiloc}}\cong
\langle \psi_{\rm D} | \hat{V}_{\text{x}}^{\text{hyb}}- \hat{V}_{\text{x}}^{\text{semiloc}}| \psi_{\rm D} \rangle,
\label{JS2}
\end{equation} 
where the exchange potential $\hat{V}_{\text{x}}$ is evaluated at half filling.
Two different contributions can conceptually be distinguished in Eq.\ (\ref{JS2}) referring to 
defect-defect and defect-bulk interactions. The separation is trivial for the Hartree-Fock 
exchange term, but requires some prescription for semilocal exchange. 
We first focus on defect-defect contributions. Since the defect wave function $\psi_{\rm D}$ 
is atomically localized, differences due to these interactions between charge transitions levels 
derived in the two schemes should be analogous to the corresponding differences for ionization potentials and 
electron affinities of atoms and molecules.  The latter quantities can be expressed as total-energy 
differences and are already well described in semilocal approximations \cite{Ogut_PRL_1998,Perdew_PRB_1997}, 
as demonstrated by extensive quantum chemistry calculations \cite{Curtiss_JCP_1997}. Hence, this 
contribution is expected to give energy-level differences independent of the location of the defect level 
in the band gap.  As for the defect-bulk contributions, it can be shown that they vanish in the 
limit of point-like defect states. When the defect wave function has a finite extent, 
these contributions depend on the degree of valence-band vs.\ conduction-band character 
of the defect state and can lead to a slope larger than 1 in Fig.\ \ref{fig2}.
To support this picture, we calculated average spreads of the defect wave functions 
in each material. Figure \ref{fig3}(b) clearly shows that the slopes of the linear 
regressions increase with these spreads, as the band gap decreases.

%
Our results reveal a general trend which appears amenable to generalization. 
When the theoretical description is improved, band edges in these materials undergo significant shifts 
but charge transition levels of atomically localized defects remain practically unaffected.  
This leads us to propose the following guideline to locate charge transition levels in the 
experimental band gap.  First, ordinary semilocal density-functional calculations are performed and
charge transition levels of the targeted defect as well as band edges are determined.
Second, the positions of the band edges are corrected through the use of a high-level electronic-structure theory
which yields a band gap in agreement with experiment, e.g.\ through hybrid density-functional or $GW$ calculations.
This only requires a calculation for the bulk material, which is computationally less demanding than a defect calculation.
Third, the defect level is located in the new band structure following the alignment proposed in this work.
We demonstrate the applicability of this scheme for well-characterized defect levels: 
the two donor levels of substitutional Te in silicon (Te$_{\rm  Si}$) \cite{Kalyanaraman_JAP_1983}, the acceptor level 
of interstitial C in silicon (C$_{\rm i}$) \cite{Song_PRB_1990}, and the optical transition
between the valence band and the $E_1^\prime$ defect state in $\alpha$-quartz \cite{Skuja_JNCS_1998,Pacchioni_PRL_1998}. 
The levels are first determined with respect to the valence band maximum within the semilocal scheme \cite{note2}. The position of the valence band is 
then adjusted according to recent $GW$ calculations \cite{Shaltaf_PRL_2008}. The resulting defect levels agree with the 
measured ones within the errors expected from our analysis (Table \ref{tab2}).

\begin{table}
\caption{Comparison between calculated ($\mu^\text{th}$) and measured ($\mu^\text{expt}$) defect levels given with respect to
the valence band. Theoretical levels are obtained from semilocal levels ($\mu^\text{semiloc}$) through application of $GW$ corrections 
to the valence band edge ($\Delta\varepsilon_\text{v}^{GW}$) \cite{Shaltaf_PRL_2008}. Experimental data for the defects 
Te$_{\rm  Si}$, C$_{\rm i}$, and $E_1^\prime$ are from Refs.\ \cite{Kalyanaraman_JAP_1983}, 
\cite{Song_PRB_1990}, and \cite{Skuja_JNCS_1998}, respectively.}
\begin{ruledtabular}
\begin{tabular}{lcccccc}
       &  Defect & $q/q^\prime$   & $\mu^{\rm semiloc}$ & $\Delta\varepsilon_{\rm v}^{GW}$ & $\mu^{\rm th}$ & $\mu^{\rm expt}$   \\ \hline
Si     &  Te$_{\rm  Si}$ & 0/+    &   0.2           & $-$0.4                     & 0.6    & 0.6    \\
Si     &  Te$_{\rm  Si}$ & +/++   &   0.4           & $-$0.4                     & 0.8    & 1.0    \\
Si     &  C$_{\rm i}$    & 0/$-$  &   0.5           & $-$0.4                     & 0.9    & 1.0    \\
SiO$_2$&  $E_1^\prime$   & +/0    &   4.1           & $-$1.9                     & 6.0    & 6.0    \\ 
\end{tabular}
\label{tab2}
\end{ruledtabular}
\end{table}

This guideline applies to atomically localized defect states and is clearly inappropriate for 
effective-mass-like defect levels which are tied to band edges. Application of this procedure also requires
that the defect is well described already within the semilocal density-functional scheme.  For instance, the
defect level should fall within the reduced band gap of the latter scheme to preserve its localized nature.
An inaccurate description may also result from the occurrence of competition between defect states
featuring different degrees of localization \cite{Laegsgaard_PRL_2001}.

%
Our findings relate to other studies of defect levels and band gaps.
Indeed, the band gap can also change as a result of a physical process,
such as quantum confinement. For quantum dots of varying size, it has been shown that
ionization potentials of deep defects remain constant as the band gap changes \cite{Melnikov_PRL_2004}.
These potentials correspond to charge transition levels referred to the vacuum level.
Another way to modify the band gap is achieved by changing the host material. It has been found
that energy levels of transition-metal impurities within a set of isovalent semiconductors are aligned when
referred to the vacuum level \cite{Caldas_APL_1984}. From the perspective of the
present work, such an alignment is understood to the extent that the local chemistry 
of the defect is preserved and a common reference potential can be identified. 
Such transition-metal markers can then be used to predict band-offsets at interfaces 
\cite{Langer_PRL_1985}.

In conclusion, calculated energy levels of atomically localized defects generally remain 
tied to a suitably defined reference level as the description of the band gap is improved. This leads to 
a guideline for comparing calculated and measured defect levels even when the adopted 
theoretical scheme is subject to the band gap problem.

We thank A.~Baldereschi, S.~de Gironcoli, and J.~Hutter for fruitful interactions.
Support from the Swiss National Science Foundation (Grant No.\ 200020-111747) is acknowledged.
The calculations were performed on the BlueGene of EPFL, and at DIT-EPFL and CSCS.


\begin{thebibliography}{99}
\bibitem{VanDeWalle_JAP_2004} C. G. Van de Walle and J. Neugebauer, J. Appl. Phys. \textbf{95}, 3851 (2004).

\bibitem{Lany_PRL_2007} S. Lany and A. Zunger, Phys. Rev. Lett. \textbf{98}, 045501 (2007).

\bibitem{Ismail-Beigi_PRL_2005} S. Ismail-Beigi and S. G. Louie, Phys. Rev. Lett. \textbf{95}, 156401 (2005);
M. Hedstr\"{o}m \emph{et al.}, \emph{ibid.} \textbf{97}, 226401 (2006).

\bibitem{marker}
J. Coutinho \emph{et al.},
Phys. Rev. B \textbf{73}, 235213 (2006).

\bibitem{Janotti_PRB_2007} A. Janotti and C. G. Van de Walle, Phys. Rev. B \textbf{76}, 165202 (2007).

\bibitem{Stampfl_PRB_2000} C. Stampfl \emph{et al.}, Phys. Rev. B \textbf{61}, R7846 (2000);
J. Li and S.-H. Wei, Phys. Rev. B \textbf{73}, 041201(R) (2006).

\bibitem{Zhang_PRB_2001} S. B. Zhang, S.-H. Wei, and A. Zunger, Phys. Rev. B \textbf{63}, 075205 (2001).

\bibitem{Knaup_PRB_2005}
J. M. Knaup \emph{et al.}, Phys. Rev. B \textbf{72}, 115323 (2005);
K. Xiong \emph{et al.}, Appl. Phys. Lett. \textbf{87}, 183505 (2005);
J. L. Gavartin \emph{et al.}, \emph{ibid.} \textbf{89}, 082908 (2006);
P. Broqvist and A. Pasquarello, \emph{ibid.} \textbf{89}, 262904 (2006).

\bibitem{Curtiss_JCP_1997} L.A. Curtiss \emph{et al.}, J. Chem. Phys. 
\textbf{109}, 42 (1998).

\bibitem{Muscat_CPL_2001}
J. Muscat, A. Wander, and N. M. Harrison, Chem. Phys. Lett. \textbf{342}, 397 (2001);
J. Paier \emph{et al.}, J. Chem. Phys. \textbf{124}, 154709 (2006); \emph{ibid.} \textbf{125},
249901 (2006).

\bibitem{PBE_PRL_1996} J. P. Perdew, K. Burke, and M. Ernzerhof, Phys. Rev. Lett. \textbf{77}, 3865 (1996).
\bibitem{PBE0} J. P. Perdew, M. Ernzerhof, and K. Burke, J. Chem. Phys. \textbf{105}, 9982 (1996).


\bibitem{Gygi_PRB_1986}
F. Gygi and A. Baldereschi, Phys. Rev. B \textbf{34}, 4405 (1986).

\bibitem{note1} Relaxations at the hybrid level for selected defects yielded negligible differences.

\bibitem{QE} http://www.quantum-espresso.org.

\bibitem{CPMD}
{\sc cpmd}, Copyright IBM Corp 1990-2006, Copyright MPI f\"{u}r 
Festk\"{o}rperforsch.\ Stuttgart 1997-2001.

\bibitem{Schultz_PRL_2006} P. A. Schultz, Phys.\ Rev.\ Lett.\ \textbf{96}, 246401 (2006).

\bibitem{Alkauskas_PB_2007} A. Alkauskas and A. Pasquarello, Physica B \textbf{401-402}, 670 (2007). 

\bibitem{Slater} J. C. Slater, Adv. Quantum Chem. \textbf{6}, 1 (1972).


\bibitem{Ogut_PRL_1998}
S. \"{O}\v{g}\"{u}t, J. R. Chelikowsky, and S. G. Louie,  Phys. Rev. Lett. \textbf{80}, 3162 (1998).

\bibitem{Perdew_PRB_1997} J. P. Perdew and M. Levy, Phys. Rev. B \textbf{56}, 16021 (1997).

\bibitem{Kalyanaraman_JAP_1983} V. Kalyanaraman, M. M. Chandra, and V. Kumar, J. Appl. Phys. \textbf{54}, 6417 (1983).

\bibitem{Song_PRB_1990} L. W. Song and G. D. Watkins, Phys. Rev. B \textbf{42}, 5759 (1990).

\bibitem{Skuja_JNCS_1998} L. Skuja, J. Non-Cryst. Solids \textbf{239}, 16 (1998).

\bibitem{Pacchioni_PRL_1998}
G. Pacchioni, G. Ieran\'{o}, and A. M. M\'{a}rquez, Phys. Rev. Lett. \textbf{81}, 377 (1998).

\bibitem{note2}
For Si, the reported defect levels are extrapolated 
from supercell calculations with 64, 216 and 512 atoms. 

\bibitem{Shaltaf_PRL_2008} R. Shaltaf {\it et al.}, 
Phys.\ Rev.\ Lett.\ {\bf 100}, 186401 (2008).

\bibitem{Laegsgaard_PRL_2001}
G. Pacchioni \emph{et al.}, Phys. Rev. B {\bf 63}, 054102 (2000).

\bibitem{Melnikov_PRL_2004} D. V. Melnikov and J. R. Chelikowsky, Phys. Rev. Lett. \textbf{92}, 046802 (2004).

\bibitem{Caldas_APL_1984} 
M. J. Caldas, A. Fazzio, and A. Zunger, Appl. Phys. Lett. \textbf{45}, 671 (1984).

\bibitem{Langer_PRL_1985}
J. M. Langer and H. Heinrich, Phys.\ Rev.\ Lett.\ {\bf 55}, 1414 (1985).


\end{thebibliography}
\end{document}